\documentclass[12pt]{iopart}
\usepackage{epsf}
\newcommand{\be}{\begin{equation}}
\newcommand{\ee}{\end{equation}}
\newcommand{\bea}{\begin{eqnarray}}
\newcommand{\eea}{\end{eqnarray}}

\begin{document}

\title[Scattering mechanisms and spectral properties...]
{Scattering mechanisms and spectral properties
of the one-dimensional Hubbard model}
\author{J. M. P. Carmelo}
\address{Department of Physics, Massachusetts Institute of
Technology, Cambridge, Massachusetts 02139-4307, USA and\\
GCEP and Department of Physics,
University of Minho, Campus Gualtar, P-4710-057 Braga, Portugal}
\date{5 May 2005}


\begin{abstract}
It is found that the finite-energy spectral properties of the one-dimensional Hubbard
model are controlled by the scattering of charged $\eta$-spin-zero $2\nu$-holon composite
objects, spin-zero $2\nu$-spinon composite objects, and charged $\eta$-spin-less and
spin-less objects, rather than by the scattering of independent $\eta$-spin $1/2$ holons
and spin $1/2$ spinons. Here $\nu =1,2,...$. The corresponding $S$ matrix is calculated
and its relation to the spectral properties is clarified.
\end{abstract}


\maketitle

The description of the microscopic scattering mechanisms behind the unusual finite-energy
spectral properties observed in low-dimensional materials remains until now an
interesting open problem. The one-dimensional (1D) Hubbard Hamiltonian is the simplest
model for the description of electronic correlations in a chain of $N_a$ sites. It reads
$\hat{H} = \hat{T}+U\,\hat{D} - [U/2][\hat{N} - N_a/2]$, where
$\hat{T}=-t\sum_{\sigma=\uparrow ,\,\downarrow
}\sum_{j=1}^{N_a}[c_{j,\,\sigma}^{\dag}\,c_{j+1,\,\sigma} + h. c.]$ is the {\it
kinetic-energy} operator, $\hat{D} =
\sum_{j}\hat{n}_{j,\,\uparrow}\,\hat{n}_{j,\,\downarrow}$ the electron double-occupation
operator, $\hat{N}= \sum_{j,\,\sigma}\hat{n}_{j,\,\sigma}$ the electron number operator,
and the operator $c_{j,\,\sigma}^{\dag}$ creates a spin-$\sigma$ electron at site $j$. In
contrast to other interacting models \cite{Lee} and in spite of the model exact solution
\cite{Lieb}, until recently little was known about its finite-energy spectral properties
for finite values of the on-site repulsion $U$. Recently, the problem was studied by the
pseudofermion dynamical theory (PDT) introduced in Refs. \cite{IIIb,V}, whose predictions
agree quantitatively for the whole momentum and energy bandwidth with the peak
dispersions observed for the TCNQ stacks by angle-resolved photoelectron spectroscopy in
the quasi-1D conductor TTF-TCNQ and are consistent with the phase diagram observed for
the $\rm{(TMTTF)_2X}$ and $\rm{(TMTSF)_2X}$ series of compounds \cite{Applications}. More
recently, results for the TTF-TCNQ spectrum consistent with those of the PDT were
obtained by the dynamical density matrix renormalization group method \cite{Eric}. Within
the PDT, the finite-energy spectral properties are controlled by the functional character
of the pseudofermion anticommutators \cite{V,Applications}. However, the relation of
these anticommutators to the elementary-excitation $S$ matrix remains an open question.
Moreover, the fact that these anticommutators do not couple quantum objects with
different $\eta$-spin or spin projections seems to be inconsistent with the form of the
$S$ matrix for elementary excitations calculated in Refs. \cite{Natan,S}. Thus, the study
of the relation of the PDT to the elementary-excitation scattering is an important issue
both for the clarification of that apparent inconsistency and the further understanding
of the scattering mechanisms that control the exotic finite-energy spectral properties of
low-dimensional materials and of the new quantum systems described by cold fermionic
atoms on an optical lattice \cite{Applications,Jaksch}.

In this Letter the above problems are solved by identifying the active scatterers and
scattering centers which control the dynamical properties, calculating their $S$ matrix,
and clarifying its relation to the spectral properties. Moreover, the connection to the
$S$ matrix of Refs. \cite{Natan,S} is also clarified. The number of lattice sites $N_a$
is considered large, units of Planck constant and lattice spacing one are used, and the
lattice length is denoted by $L=N_a$ and the electronic charge by $-e$. The densities
$n=N/L$ and spin densities $m=[N_{\uparrow}-N_{\downarrow}]/L$ are in the domains $0<
n\leq 1$ and $0\leq m < n$, respectively. The above Hamiltonian commutes with the
generators of the $\eta$-spin and spin $SU(2)$ algebras \cite{I}. Here the $\eta$-spin
and spin values of an energy eigenstate are called $\eta$ and $S$, respectively, and the
corresponding projections $\eta_z$ and $S_z$. A key result needed for our study is that
all energy eigenstates of the model can be described in terms of occupancy configurations
of $\eta$-spin $1/2$ holons, spin $1/2$, spinons, and $\eta$-spin-less and spin-less $c0$
pseudoparticles \cite{I}. Below, the notation $\pm 1/2$ holons and $\pm 1/2$ spinons is
used according to the values of $\eta$-spin and spin projections, respectively. The
electron - rotated-electron unitary transformation \cite{I} maps the electrons onto
rotated electrons such that rotated-electron double occupation, no occupation, and
spin-up and spin-down single occupation are good quantum numbers for all values of $U$.
The $\pm 1/2$ holons of charge $\pm 2e$ and zero spin and the charge-less $\pm 1/2$
spinons are generated from the electrons by that unitary transformation. The
corresponding holon and spinon number operators ${\hat{M}}_{c,\,\pm 1/2}$ and
${\hat{M}}_{s,\,\pm 1/2}$, respectively, are of the form given in Eq. (24) of \cite{I}
and involve the electron - rotated-electron unitary operator. While the $-1/2$ and $+1/2$
holons refer to the rotated-electron doubly occupied and unoccupied sites, respectively,
the $-1/2$ and $+1/2$ spinons correspond to the spin degrees of freedom of the spin-down
and spin-up rotated-electron singly occupied sites, respectively. The charge degrees of
freedom of the latter sites are described by the spin-less and $\eta$-spin-less $c0$
pseudoparticles, which are composite objects of a charge $-e$ chargeon and a charge $+e$
antichargeon \cite{I}. The $c\nu$ pseudoparticles (and $s\nu$ pseudoparticles) such that
$\nu =1,2,...$ are $\eta$-spin singlet (and spin singlet) $2\nu$-holon (and
$2\nu$-spinon) composite objects. Thus, $M_{\alpha,\,\pm 1/2}=L_{\alpha,\,\pm 1/2}
+\sum_{\nu =1}^{\infty}\nu\,N_{\alpha\nu}$ where $\alpha =c,\,s$, $N_{\alpha\nu}$ denotes
the number of $\alpha\nu$ pseudoparticles, and $L_{c,\,\pm 1/2}=\eta\mp \eta_z$ and
$L_{s,\,\pm 1/2}=S\mp S_z$ gives the number of $\pm 1/2$ Yang holons and $\pm 1/2$ HL
spinons, respectively. Those are the holons and spinons that are not part of composite
pseudoparticles. All energy eigenstates can be described by occupancy configurations of
$\alpha\nu$ pseudoparticles, $-1/2$ Yang holons, and $-1/2$ HL spinons \cite{I}. For the
ground state, $N_{c0}=N$, $N_{s1}=N_{\downarrow}$,
$N_{\alpha\nu}=L_{c,\,-1/2}=L_{s,\,-1/2}=0$ for $\alpha\nu\neq c0,\,s1$.

In our study we consider the {\it pseudofermion subspace} (PS),
which is spanned by the initial ground state $\vert GS\rangle$ and
all excited energy eigenstates contained in $\hat{O}\vert
GS\rangle$, where $\hat{O}$ is any one-electron or two-electron
operator. In reference \cite{IIIb} it is shown that within the PS
there is a unitary transformation that maps the $\alpha\nu$ {\it
pseudoparticle} or hole onto the $\alpha\nu$ {\it pseudofermion}
or hole, respectively. These objects differ only in the discrete
momentum values. The $\alpha\nu$ pseudoparticle or hole has
discrete bare-momentum values $q_j=[2\pi/L]I^{\alpha\nu}_j$ such
that $I^{\alpha\nu}_j$ are consecutive integers or half-odd
integers \cite{I}. These values are good quantum numbers whose
allowed occupancies are one (pseudoparticle) and zero (hole) only.
The $\alpha\nu$ pseudofermion or hole has discrete
canonical-momentum values given by,

\begin{equation}
{\bar{q}}_j = {\bar{q}} (q_j) = q_j + Q^{\Phi}_{\alpha\nu} (q_j)/L
\, , \label{barqan}
\end{equation}
where $j=1,2,...,N_{\alpha\nu}^*$, $N^*_{\alpha\nu}=N_{\alpha\nu}+N_{\alpha\nu}^h$, and
$N_{\alpha\nu}^h$ denotes the number of $\alpha\nu$ pseudofermion holes, which equals
that of $\alpha\nu$ pseudoparticle holes, whose value is given in Eq. (B.11) of \cite{I}.
Such a canonical-momentum pseudofermion is related in \cite{IIIb} to the {\it local
$\alpha\nu$ pseudofermion} by a suitable Fourier transformation. The latter object
occupies the sites of the effective $\alpha\nu$ lattice \cite{IIIb,V}. Except for the
discrete momentum values, the above pseudoparticle and pseudofermion have the same
properties. Thus, all the energy eigenstates that span the PS can be described by
occupancy configurations of $\alpha\nu$ pseudofermions, $-1/2$ Yang holons, and $-1/2$ HL
spinons \cite{IIIb,V}. The functional,
\begin{equation}
Q^{\Phi}_{\alpha\nu} (q_j) = 2\pi \sum_{\alpha'\nu',\,j}
\Phi_{\alpha\nu,\,\alpha'\nu'}(q_j,q_{j'})\, \Delta
N_{\alpha'\nu'}(q_{j'}) \, , \label{qcan1j}
\end{equation}
of equation (\ref{barqan}) was introduced in \cite{IIIb} and is such that
$Q^{\Phi}_{\alpha\nu} (q_j)/2$ is found below to be an overall scattering phase shift.
Here $\Delta N_{\alpha\nu} (q_j) \equiv N_{\alpha\nu} (q_j) - N^{0}_{\alpha\nu} (q_j)$ is
the $\alpha\nu$ branch bare-momentum distribution-function deviation relative to the
ground state value and $\pi\,\Phi_{\alpha\nu,\,\alpha'\nu'}(q,q')$ is defined in
\cite{IIIb} and is found below to be an elementary {\it two-pseudofermion phase shift}.
Note that $Q^{\Phi}_{\alpha\nu} (q_j)=0$ for the initial ground state and thus
${\bar{q}}_j= q_j$ for that state.

Each transition from the initial ground state to a PS excited energy eigenstate can be
divided into two elementary processes. The first process is a scattering-less
finite-energy and finite-momentum excitation which transforms the ground state onto a
well defined virtual state. This excitation involves the pseudofermion creation,
annihilation, and particle-hole processes associated with the PS excited state and the
discrete bare-momentum shift $Q_{\alpha\nu}^0/L$, whose possible values are
$0,\,\pm\pi/L$ \cite{IIIb}, for $\alpha\nu$ branches with finite occupancy in that state.
For $\nu>0$ branches that excitation can involve a change in the number of discrete
bare-momentum values. Although the $\alpha\nu\neq c0,\, s1$ branches have no finite
pseudofermion occupancy in the initial ground state, one can define the values
$N^*_{\alpha\nu}=N^h_{\alpha\nu}$ for the corresponding empty bands \cite{I,IIIb,V}. In
this first step the pseudofermions acquire the excitation momentum and energy needed for
the second-step scattering events. Thus, the virtual state is the in asymptote of the
pseudofermion scattering theory. The second elementary step of the ground-state
transition involves a set of elementary scattering events where all $\alpha\nu$
pseudofermions or holes of momentum $q_j+Q_{\alpha\nu}^0/L$ of the in asymptote are the
scatterers. Each of these elementary scattering events leads to a phase factor in the
wave function of the $\alpha\nu$ pseudofermions or holes given by,

\begin{equation}
S_{\alpha\nu ,\,\alpha'\nu'} (q_j, q_{j'}) =
e^{i2\pi\,\Phi_{\alpha\nu,\,\alpha'\nu'}(q_j,q_{j'})\, \Delta
N_{\alpha'\nu'}(q_{j'})}  \, . \label{Sanan}
\end{equation}
The scattering centers are the $\alpha'\nu'$ pseudofermions or holes of momentum
$q_{j'}+Q_{\alpha\nu}^0/L$ created in the ground-state - virtual-state transition and
thus such that $\Delta N_{\alpha'\nu'}(q_{j'})\neq 0$. Indeed, note that $S_{\alpha\nu
,\,\alpha'\nu'} (q_j, q_{j'})=1$ for $\Delta N_{\alpha'\nu'}(q_{j'}) =0$. There is a
one-to-one correspondence between the local rotated-electron occupancy configurations
that describe the PS energy eigenstates and the local $\alpha\nu$ pseudofermion occupancy
configurations and $-1/2$ Yang holon and $-1/2$ HL spinon occupancies that describe the
same states \cite{V}. The corresponding effective $\alpha\nu$ lattices have the same
length $L$ as the original lattice. Our analysis refers to periodic boundary conditions
and the thermodynamic limit $L\rightarrow\infty$. Under a ground-state -
excited-energy-eigenstate transition, by moving the $\alpha\nu$ pseudofermion or hole of
initial ground-state momentum $q_j$ once around the length $L$ lattice ring, its wave
function acquires the following overall phase factor,

\begin{eqnarray}
S_{\alpha\nu} (q_j) & = & e^{i\,Q_{\alpha\nu}^0}
\prod_{\alpha'\nu'}\,
\prod_{j'=1}^{N^*_{\alpha'\nu'}}\,S_{\alpha\nu ,\,\alpha'\nu'}
(q_j, q_{j'}) \nonumber \\
& = & e^{i\,Q_{\alpha\nu}(q_j)}  \, ; \hspace{0.5cm} j=1,2,...,
N^*_{\alpha\nu} \, . \label{San}
\end{eqnarray}
Interestingly, $Q_{\alpha\nu}(q_j)/L$ is the net $\alpha\nu$ pseudofermion or hole
discrete canonical-momentum shift that arises due to the above transition \cite{IIIb,V}
and thus in this equation,

\begin{equation}
Q_{\alpha\nu}(q_j) = Q_{\alpha\nu}^0 + Q^{\Phi}_{\alpha\nu} (q_j)
\, , \label{Qcan1j}
\end{equation}
is such that $Q_{\alpha\nu}(q_j)/2$ is a $\alpha\nu$ pseudofermion or hole overall phase
shift. Indeed, if when moving around the lattice ring the $\alpha\nu$ pseudofermion or
hole departures from the point $x=-L/2$ and arrives to $x=L/2$, one finds that
$\lim_{x\rightarrow L/2}\,\bar{q}\,x = q\,x + Q_{\alpha\nu} (q)/2$ where $q$ refers to
the initial ground state. From Eqs. (\ref{qcan1j}) and (\ref{Qcan1j}) it then follows
that $\pi\,\Phi_{\alpha\nu,\,\alpha'\nu'}(q_j,q_{j'})$ is an elementary two-pseudofermion
phase shift. (If instead one considers $x=0$ and $x=L$, the overall phase shift and the
two-pseudofermion phase shifts read $Q_{\alpha\nu} (q)$ and
$2\pi\,\Phi_{\alpha\nu,\,\alpha'\nu'}(q_j,q_{j'})$, respectively \cite{details}. However,
the choice of either definition is a matter of taste and the uniquely defined quantity is
the above $S$ matrix.)

Several properties play an important role in the pseudofermion scattering theory. First,
the elementary scattering processes associated with the phase factor (\ref{Sanan})
conserve the total energy and total momentum. Second, the elementary scattering processes
are of forward-scattering type and thus conserve the individual in asymptote $\alpha\nu$
pseudofermion or hole momentum and energy. These processes also conserve the $\alpha\nu$
branch, usually called {\it channel} in the scattering language. Moreover, the scattering
amplitude does not connect objects with different $\eta$ spin or spin. Last but not
least, for each $\alpha\nu$ pseudofermion or hole of initial ground-state momentum $q_j$,
the $S$ matrix associated with the ground-state - excited-energy-eigenstate transition is
simply the phase factor given in Eq. (\ref{San}). For each excited energy eigenstate (out
asymptote) the number of $\alpha\nu$ pseudofermions plus the number of $\alpha\nu$
pseudofermion holes whose $S$ matrix is of the form (\ref{San}) is given by $N_a +
N^*_{s1} + \sum_{\alpha\nu\neq c0 ,\,s1} \theta (\vert\Delta N_{\alpha\nu}\vert)\,
N^*_{\alpha\nu}$. Here $\theta (x)=1$ for $x>0$ and $\theta (x)=0$ for $x= 0$.

Importantly, the form of the scattering part of the overall phase
shift (\ref{Qcan1j}), Eq. (\ref{qcan1j}), reveals that the value
of such a phase-shift functional is independent of the changes in
the occupation numbers of the $\pm 1/2$ Yang holons and $\pm 1/2$
HL spinons. Thus, these objects are not scattering centers.
Moreover, they are not scatterers, once their momentum values
remain unchanged under the ground-state -
excited-energy-eigenstate transitions. In turn, the pseudofermions
and holes are scatterers and scattering centers. Since the $c0$
pseudofermion is a $\eta$-spin-less and spin-less object and for
$\nu >0$ the $\alpha\nu$ pseudofermions are $\eta$-spin $(\alpha
=c)$ and spin $(\alpha =s)$ singlet $2\nu$-holon and $2\nu$-spinon
composite objects, respectively, their $S$ matrix has dimension
one: it is the phase factor (\ref{San}). The factorization of the
Bethe-ansatz (BA) bare $S$ matrix for the original spin $1/2$
electrons is associated with the so called Yang-Baxter Equation
(YBE) \cite{Natan}. On the other hand, the factorization of the
$S$ matrix (\ref{San}) in terms of the elementary $S$ matrices
$S_{\alpha\nu ,\,\alpha'\nu'} (q_j, q_{j'})$, Eq. (\ref{Sanan}),
is commutative. Such a commutativity is stronger than the symmetry
associated with the YBE and results from the elementary $S$
matrices $S_{\alpha\nu ,\,\alpha'\nu'} (q_j, q_{j'})$ being simple
phase factors, instead of matrices of dimension larger than one.
This seems to be inconsistent with all PS energy eigenstates being
described by occupancy configurations which, besides $c0$
pseudofermions, involve finite spin $1/2$ spinons and $\eta$-spin
$1/2$ holons \cite{I}. Indeed, the $S$ matrix of finite
$\eta$-spin or spin objects has dimension larger than one.
However, due to the correlations the quantum liquid self organizes
in such a way that the scatterers and scattering centers are the
$c0$ pseudofermions, $\eta$-spin singlet $2\nu$-holon composite
$c\nu$ pseudofermions, and spin singlet $2\nu$-spinon composite
$s\nu$ pseudofermions.

Let us clarify how the $\alpha\nu$ pseudofermion $S$ matrix (\ref{San}) controls the
unusual spectral properties of the model. Consider a $\alpha\nu$ pseudofermion of
canonical momentum ${\bar{q}}$ and a $\alpha'\nu'$ pseudofermion of canonical momentum
${\bar{q}'}$ such that the values ${\bar{q}}$ and ${\bar{q}'}$ correspond to a PS excited
energy eigenstate and the initial ground state, respectively, and thus ${\bar{q}'}=q'$.
Importantly, from the use of Eq. (\ref{San}) it is found that the pseudofermion
anticommutation relations introduced in \cite{IIIb} can be expressed solely in terms of
the difference $[{\bar{q}}-{\bar{q}}']$ and the $S$ matrix of the $\alpha\nu$
pseudofermion associated with the excited state,

\begin{eqnarray}
& & \{f^{\dag
}_{{\bar{q}},\,\alpha\nu},\,f_{{\bar{q}}',\,\alpha'\nu'}\} =
\delta_{\alpha\nu,\,\alpha'\nu'} \nonumber
\\
& & \times\,{1\over N^*_{\alpha\nu}}\,\Bigl[S_{\alpha\nu}
(q)\Bigr]^{1/2}\,e^{-i({\bar{q}}-{\bar{q}}')/ 2}\,{{
Im}\Bigl[S_{\alpha\nu} (q)\Bigr]^{1/2}\over\sin
([{\bar{q}}-{\bar{q}}']/2)} \, , \label{pfacrGS-S}
\end{eqnarray}
and the anticommutators between two creation or annihilation operators vanish. This
reveals that the $S$ matrix (\ref{San}) fully controls the pseudofermion anticommutators.
Since within the PDT these anticommutators determine the value of the matrix elements
between energy eigenstates \cite{V}, it follows that the $S$ matrix (\ref{San}) controls
the spectral properties. If it had dimension larger than one, the pseudofermion algebra
would be much more involved, for the pseudofermion anticommutators would also be matrices
of dimension larger than one.

In reference \cite{S} the excited states generated from the $n=1$ and $m=0$ ground state
were described in terms of $\pm 1/2$ holon and $\pm 1/2$ spinon occupancy configurations.
Following the analysis of Refs. \cite{Faddeev,Faddeev-2} for the related spin $1/2$
isotropic Heinsenberg chain, the holes of the BA length-one spin string spectrum (spin
singlet two-spinon composite $s1$ pseudoparticle spectrum) were identified in \cite{S}
with the spinons. Inspired in such an interpretation, the studies of the latter reference
identified the holons with the holes of the BA distribution of $k's$ \cite{Lieb,I}
spectrum [$c0$ pseudoparticle spectrum]. This is behind the charge $\pm e$ found for the
$\pm 1/2$ holons in \cite{S}, which is half of the value found in \cite{I}. However, the
$c0$ pseudoparticle and hole band occupancy configurations do not correspond to
$\eta$-spin $SU(2)$ irreducible representations. Indeed, in \cite{I} it is shown that for
the whole Hilbert space all such representations exactly correspond to the BA charge
string and $\mp 1/2$ Yang holon occupancy configurations. Following directly the analysis
of Refs. \cite{Faddeev,Faddeev-2}, the studies of \cite{S} consider that the $\pm 1/2$
holons and $\pm 1/2$ spinons are the scatterers and scattering centers. This leads to two
$4\times 4$ $S$ matrices for holons and spinons, respectively, and a related $16\times
16$ $S$ matrix for the full scattering problem. In spite of being mathematically elegant
and obeying the YBE, these matrices are not suitable for the description of the spectral
properties. Moreover, provided that within the $x=0$ and $x=L$ boundary conditions one
defines the overall phase shift as $Q_{\alpha\nu}(q)$, the phase shifts given in Eqs.
(5.19)-(5.21) of the phase shifts given in Eqs. (5.19)-(5.21) of the first paper of
\cite{S}, which appear in the entries of these matrices, are nothing but very particular
cases of $\eta$-spin-less and spin-less $c0$ pseudofermion hole or spin-zero $s1$
pseudofermion hole overall phase shifts given in Eq. (\ref{Qcan1j}) \cite{details}.
Indeed, these phase shifts correspond to the $n=1$ and $m=0$ initial ground state and the
specific excited states considered in Refs. \cite{Natan,S}. Let $q_1$ or ${q'}_1$ be the
bare momenta of the scattered $c0$ or $s1$ pseudofermion hole, respectively, of the
latter states. For the $\eta$-spin triplet, $\eta$-spin singlet, and $\eta$-spin and spin
doublet excited states considered in these references, it is found that $\pi + Q_{c0}
(q_1)$ equals the phase shift $\delta_{CT}$ and $\delta_{CS}$ given in Eq. (5.19) and
$\delta_{\eta S}$ in Eq. (5.21) of the above paper, respectively. For the spin triplet,
spin singlet, and spin and $\eta$-spin doublet excited states, $Q_{s1} ({q'}_1)$ equals
the shift functions $\delta_{ST}$ and $\delta_{SS}$ given in Eq. (5.20) and $\delta_{S
\eta}$ given in Eq. (5.21), respectively \cite{details}. Thus, the BA phase shifts of
Refs. \cite{Natan,S} are particular cases of the $c0$ and $s1$ pseudofermion hole overall
phase-shift functionals of Eq. (\ref{Qcan1j}) and are associated with a set of excited
states which span a subspace smaller than the PS of the one- and two-electron
excitations. According to the studies of \cite{I}, for all the transitions associated
with these phase shifts the deviations in the $\eta$-spin and spin values are provided by
the $\pm 1/2$ Yang holons and $\pm 1/2$ HL spinons occupancy changes, respectively, which
do not contribute to the phase-shift values. In turn, the holes created in the $c0$ and
$s1$ bands by these transitions are both scatterers and scattering centers and it follows
from the analysis of \cite{I} that they do not correspond to single $\pm 1/2$ holons and
$\pm 1/2$ spinons, respectively. Moreover, the phase shifts of Refs. \cite{Natan,S} were
evaluated up to an overall constant term by the method of \cite{Koprepin79}. Equation
(\ref{Qcan1j}) provides the full phase shift value and reveals that the above extra $\pi$
in $\delta_{CT}$, $\delta_{CS}$, and $\delta_{\eta S}$ is not physical, as discussed
elsewhere \cite{details}. The results reported here also apply to other models. For
instance, for the isotropic Heisenberg chain it is found that the phase changes for the
spin singlet and triplet excited states given in Eq. (11) of \cite{Faddeev} equal the
phase shifts for the same states of a scattered hole of the zero-spin two-spinon $s1$
pseudofermion spectrum. Thus, for the study of the spectral properties, these two excited
states correspond to two $s1$ pseudofermion hole $S$ matrices, rather than to the single
$4\times 4$ $S$ matrix of Eq. (5.1) of \cite{Faddeev-2}. This analysis can also be
extended to the same model with an odd number of lattice sites.

While, through the anticommutators (\ref{pfacrGS-S}), the use of the $S$-matrix
introduced here leads to a successful description of the spectral features observed in
real materials \cite{Applications}, the $16\times 16$ $S$ matrix of \cite{S} is
unsuitable for such a task. Indeed, independent $\pm 1/2$ holons and $\pm 1/2$ spinons
that are not part of composite pseudofermions are neither scatterers nor scattering
centers. Interestingly, these objects remain invariant under the electron -
rotated-electron unitary, whereas the pseudofermion and holes are not in general
invariant under such a transformation.

The method for evaluation of the finite-energy spectral-weight distributions of a 1D
correlated metal introduced in \cite{V} fully relies on the scattering theory introduced
here. The exotic metallic quantum phase of matter found for quasi-1D compounds
\cite{Applications} by use of such a method is expected to emerge at finite energies in
carbon nanotubes, ballistic wires, and systems of cold fermionic atoms in one-dimensional
optical lattices with on-site atomic repulsion \cite{Jaksch}. This confirms the general
scientific interest of the scattering theory introduced here. While in this Letter it is
applied specifically to the 1D Hubbard model, the theory is of general nature for many
integrable quantum problems and therefore will have wide applicability.

I thank N. Andrei, D. Bozi, A. Castro Neto, F. Guinea, V.E. Korepin, P.A. Lee, Sung-Sik
Lee, G. Ortiz, K. Penc, T. Ribeiro, and P.D. Sacramento for illuminating discussions, the
hospitality of the MIT, and the financial support of the Gulbenkian Foundation, Fulbright
Commission, and FCT grant POCTI/FIS/58133/2004.


\end{document}